\documentstyle[amssymb,preprint,aps,epsf]{revtex}
\begin{document}
\draft
\title{Conditional Spontaneous Spin Polarization and Bifurcation Delay in Coupled
Two-Component Bose-Einstein Condensates}
\author{Chaohong Lee$^{1\thanks{%
E-mail of the corresponding author: chlee@wipm.ac.cn and
chaohonglee@hotmail.com.}}$, Wenhua Hai$^{2}$, Lei Shi$^{1}$, and Kelin Gao$%
^{1}$}
\address{{\small \ }$^{1}$Laboratory of Magnetic Resonance and Atomic and Molecular
Physics, Wuhan Institute of Physics and Mathematics, The Chinese Academy of
Sciences, Wuhan, 430071, People's Republic of China\\
$^{2}$Department of Physics, Hunan Normal University, Changsha, 410081,
People's Republic of China}
\date{\today}
\maketitle

\begin{abstract}
The spontaneous spin polarization and bifurcation delay in two-component
Bose-Einstein condensates coupled with Raman pulses are investigated. We
find that the bifurcation and the spontaneous spin polarization depend not
only on the system parameters, but also on the relative phase between two
components. Through bifurcations, the system enters into the spontaneous
spin polarization regime from the Rabi regime. We also find that bifurcation
delay appears when the parameter is swept through the static bifurcation
point. This bifurcation delay is responsible for metastability leading to
hysteresis. The area enclosed in the hysteresis loop increases with the
sweeping rate of the parameter.
\end{abstract}



Electronic and nuclear -spin polarization in an atomic vapor with optical
pumping have been investigated extensively\cite{Fortson}. Under conditions
in which electronic spin exchange takes place faster than spin relaxation,
spontaneous spin polarization appears. This interesting phenomenon is very
similar to ferromagnetism and has been observed in wide ranges of atomic
intensity, pump laser frequency and intensity. The appearance of spontaneous
spin polarization means that the atomic vapor has two stable states with
large spin polarization. The experimental realization of it has been applied
to the field of optical bistability\cite{Gibbs}. The atomic spin
polarization exhibits striking hysteresis in switching between the bistable
states\cite{Fortson}. This is analogs to those ferromagnetic system displays
magnetic hysteresis\cite{Rao}.

All the previous works have only considered the case of the thermal atoms,
the experimental realization of multi-component Bose-Einstein condensates
(BECs)\cite{Stenger,Hall} causes our interest in considering the similar
behavior of the ultracold atoms. There are many differences between thermal
and cold atoms. The first one is that the collision among thermal atoms is
noncoherent. However, when the temperature is close to the critical
temperature realizing Bose-Einstein condensate (BEC) ($T\thicksim T_{BEC}$),
the collision among ultracold atoms is coherent due to the path between such
collision is smaller than the phase coherence length\cite{Burnett,Anglin}.
Does this coherent property play a role in the polarization process of the
ultracold atoms? Another difference between thermal atoms and cold ones is
that the interaction strength of cold atoms can be controlled easily\cite
{Burnett,Inouye}. In this letter, we firstly show the coherence among
ultracold atoms gives rise to the conditional spontaneous spin polarization
which is determined by both the system parameters and the relative phase.
Then, we present the bifurcation induced by tuning the $s$-wave interaction
between ultracold atoms. Lastly, we propose an experiment to confirm our
prediction.

We consider that the same kind of bosonic atoms, which are trapped in a
single-well potential, are condensed in two different hyperfine levels $|1>$
and $|2>$. Raman transitions between two hyperfine states are induced by the
laser fields with the effective Rabi-frequency $\Omega $ and a finite
detuning $\delta $. The internal Josephson effects\cite
{Stenger,Hall,Kohler,Williams,Lee,Kuang}, coherent coupling effects\cite
{coupling}, vortices\cite{Zoller} and spin waves\cite{spinwave} in such
systems have stimulated great interest of many theoretists and
experimentists. In the rotating frame, neglecting the damping and the
finite-temperature effects, the coupled two-component BECs system can be
described by a pair of coupled GPEs\cite{Williams,coupling,Zoller} 
\begin{equation}
\begin{array}{l}
i\hbar \frac{\partial \Psi _{2}(\stackrel{\rightharpoonup }{r},t)}{\partial t%
}=(H_{2}^{0}+H_{2}^{MF}-\frac{\delta }{2})\Psi _{2}(\stackrel{%
\rightharpoonup }{r},t)+\frac{\Omega }{2}\Psi _{1}(\stackrel{\rightharpoonup 
}{r},t), \\ 
i\hbar \frac{\partial \Psi _{1}(\stackrel{\rightharpoonup }{r},t)}{\partial t%
}=(H_{1}^{0}+H_{1}^{MF}+\frac{\delta }{2})\Psi _{1}(\stackrel{%
\rightharpoonup }{r},t)+\frac{\Omega }{2}\Psi _{2}(\stackrel{\rightharpoonup 
}{r},t).
\end{array}
\end{equation}
Here, the free evolution Hamiltonians $H_{i}^{0}=-\frac{\hbar
^{2}\triangledown ^{2}}{2m}+V_{i}(\stackrel{\rightharpoonup }{r})$ $(i=1,2)$
and the mean-field interaction Hamiltonians $H_{i}^{MF}=\frac{4\pi \hbar ^{2}%
}{m}(a_{ii}|\Psi _{i}(\stackrel{\rightharpoonup }{r},t)|^{2}+a_{ij}|\Psi
_{j}(\stackrel{\rightharpoonup }{r},t)|^{2})$ $(i,j=1,2$, $i\neq j)$ ( $%
a_{ij}$ is the scattering length between states $i$ and $j$ which satisfies $%
a_{ij}=a_{ji}$ ). For the cigar-shaped trap (the trap frequencies satisfying 
$\omega _{x}=\omega _{y}>>\omega _{z}$), when the coupling is weak enough $($%
i.e., the Rabi frequency satisfies $\Omega /\omega _{z}\ll 1)$, the
macroscopic wavefunctions can be written as the variational ansatz $\Psi
_{i}(\stackrel{\rightharpoonup }{r},t)=\psi _{i}(t)\Phi _{i}(\stackrel{%
\rightharpoonup }{r})$ $(i=1,2)$ with amplitudes $\psi _{i}(t)=\sqrt{N_{i}(t)%
}e^{i\alpha _{i}(t)}$ and spatial distributions $\Phi _{i}(\stackrel{%
\rightharpoonup }{r})$. The symbols $N_{i}(t)$ and $\alpha _{i}(t)$ are the
atomic population and phase of the $i$-th component, respectively. Due to
the coupling is very small, the spatial distributions vary slowly in time
and are very close to the adiabatic solutions to the time-independent
uncoupled case for GP equations $(1)$, being slaved by the populations\cite
{Williams}. Thus, the amplitudes obey the nonlinear two-mode dynamical
equations 
\begin{equation}
\begin{array}{l}
i\hbar \frac{d}{dt}\psi _{_{2}}(t)=(E_{2}^{0}-\frac{\delta }{2}+U_{22}|\psi
_{_{2}}(t)|^{2}+U_{21}|\psi _{_{1}}(t)|^{2})\psi _{_{2}}(t)+\frac{K}{2}\psi
_{_{1}}(t), \\ 
i\hbar \frac{d}{dt}\psi _{_{1}}(t)=(E_{1}^{0}+\frac{\delta }{2}+U_{11}|\psi
_{_{1}}(t)|^{2}+U_{12}|\psi _{_{2}}(t)|^{2})\psi _{_{1}}(t)+\frac{K}{2}\psi
_{_{2}}(t).
\end{array}
\end{equation}
The parameters $E_{i}^{0}=\int \Phi _{i}(\stackrel{\rightharpoonup }{r}%
)H_{i}^{0}\Phi _{i}(\stackrel{\rightharpoonup }{r})d\stackrel{%
\rightharpoonup }{r}$, $U_{ij}=\frac{4\pi \hbar ^{2}a_{ij}}{m}\int |\Phi
_{i}(\stackrel{\rightharpoonup }{r})|^{2}|\Phi _{j}(\stackrel{%
\rightharpoonup }{r})|^{2}d\stackrel{\rightharpoonup }{r}=U_{ji}$ and $%
K=\Omega \int \Phi _{1}(\stackrel{\rightharpoonup }{r})\Phi _{2}(\stackrel{%
\rightharpoonup }{r})d\stackrel{\rightharpoonup }{r}$ $(i,j=1,2)$. The terms
in $K$ describe the internal tunnelling between two BEC states, whereas the
terms in $U_{ij}$, which depend on the numbers of atoms in each BEC state,
describe the mean-field interaction between atoms. When $U_{21}$ and $\delta 
$ equals zero, these coupled equations can also describe the BECs in a
double-well potential \cite{doublewell}. Introducing the Bloch's spin
vectors 
\begin{equation}
u=\frac{\psi _{_{2}}^{*}\psi _{_{1}}+\psi _{_{2}}\psi _{_{1}}^{*}}{\psi
_{_{1}}^{*}\psi _{_{1}}+\psi _{_{2}}^{*}\psi _{_{2}}},\text{ }v=-i\frac{\psi
_{_{2}}\psi _{_{1}}^{*}-\psi _{_{2}}^{*}\psi _{_{1}}}{\psi _{_{1}}^{*}\psi
_{_{1}}+\psi _{_{2}}^{*}\psi _{_{2}}},\text{ }w=\frac{\psi _{_{2}}^{*}\psi
_{_{2}}-\psi _{_{1}}^{*}\psi _{_{1}}}{\psi _{_{1}}^{*}\psi _{_{1}}+\psi
_{_{2}}^{*}\psi _{_{2}}}.
\end{equation}
Obviously, $u^{2}+v^{2}+w^{2}=1$. When the total atomic numbers $N_{T}$ $=$ $%
N_{1}+N_{2}$ $=$ $\psi _{_{1}}^{*}\psi _{_{1}}+\psi _{_{2}}^{*}\psi _{_{2}}$
is conserved, setting the Planck constant $\hbar =1$, the Bloch's spin
vectors satisfy 
\begin{equation}
\frac{d}{dt}\left( 
\begin{array}{l}
u \\ 
v \\ 
w
\end{array}
\right) =\left( 
\begin{array}{ccc}
0 & \gamma +Gw & 0 \\ 
-(\gamma +Gw) & 0 & K \\ 
0 & -K & 0
\end{array}
\right) \left( 
\begin{array}{l}
u \\ 
v \\ 
w
\end{array}
\right) .
\end{equation}
In this Bloch's equation, the parameters satisfy $\gamma
=E_{2}^{0}-E_{1}^{0}+N_{T}(U_{22}-U_{11})/2-\delta $ and $%
G=N_{T}(U_{22}+U_{11}-2U_{12})/2$. Comparing the above equation with the one
for the linear case $(U_{ij}=0)$ of equation $(2)$, one can find that the
mean-field interaction induces a shift $(Gw)$ in the transition frequency
and this shift is proportional to the relative population $w$.

Taking $|1>$ as spin-up state and $|2>$ as spin-down state, the above
two-component BECs system can be regarded as an ensemble of quantum spin-$1/2
$ particles. Thus, the longitudinal component $w$ of the pseudospin
describes the relative population, and the transverse components $u$ and $v$
characterize the coherence. In this language, the effective Rabi frequency
causes an effective transverse magnetic field $K$ along axis-$u$, the
effective detuning induces an effective longitudinal magnetic field $\gamma $%
, and the mean-field interaction brings an effective longitudinal magnetic
field $Gw$ which depends on the longitudinal spin component.

From the definition of the Bloch's spin vectors, we know that the above
system can be described with only two independent variables. If we use the
longitudinal spin component $w$ and the relative phase $\phi =\alpha
_{2}-\alpha _{1}$ as independent variables, rescaling the time $Kt$ to $t$,
the motion equations 
\begin{equation}
\begin{array}{l}
dw/dt=-\sqrt{1-w^{2}}\sin \phi , \\ 
d\phi /dt=-\gamma /K-(G/K)w+w\cos \phi /\sqrt{1-w^{2}}.
\end{array}
\end{equation}
are equivalent to the Bloch's equation. The above equations are consistent
with those derived from the secondary quantized model\cite{Kohler}.

Below, let us analyze the stationary states of the coupled two-component
BECs system. The stationary states can be obtained from the stable fixed
points of the system. The fixed points correspond to those solutions
satisfying $dw/dt=0$ and $d\phi /dt=0$. In the region $[0,2\pi )$ of the
relative phase, we find two different modes of stationary states existing in
the system: one is the equal-phase mode with zero relative phase $(\phi =0)$%
, the other one is the anti-phase mode with $\pi $ relative phase $(\phi
=\pi )$.

The number of the fixed points and the stationary states depend on the
ratios $\gamma /K$, $G/K$ and the relative phase. For the equal-phase mode,
only a fixed point exists when $G/K\leq 1$ and this fixed point is stable.
When $G/K>1$, there are two stable fixed points and an unstable one for $%
(G/K)^{2/3}-(\gamma /K)^{2/3}>1$; there is only one stable fixed point for $%
(G/K)^{2/3}-(\gamma /K)^{2/3}<1$; the saddle-node bifurcations occur at the
points satisfying $(G/K)^{2/3}-(\gamma /K)^{2/3}=1$. In the left column of
Fig. 1, for the equal-phase mode, we show the values for the longitudinal
component of the fixed points with different ratios $\gamma /K$ and $G/K$.
For the anti-phase mode, the corresponding case is different. When $G/K\geq
-1$, only a fixed point appears and it is stable. When $G/K<-1$, two stable
fixed points and an unstable one exist for $(G/K)^{2/3}-(\gamma /K)^{2/3}>1$%
; only one stable fixed point emerges for $(G/K)^{2/3}-(\gamma /K)^{2/3}<1$;
the saddle-node bifurcations occur at the points satisfying $%
(G/K)^{2/3}-(\gamma /K)^{2/3}=1$. The fixed points of the anti-phase mode
with different ratios $\gamma /K$ and $G/K$ are exhibited in the right
column of Fig. 1. In the Fig. 1, the fixed points between a pair of
corresponding bifurcation points are unstable and the values for $d(\gamma
/K)/dw$ at the bifurcation points equal zero. From the previous analysis, we
find bistability exists in either the equal-phase mode or the anti-phase
mode when the parameters obey $(G/K)^{2/3}-(\gamma /K)^{2/3}>1$. The
appearance of bistability indicates the existence of spontaneous spin
polarization ($<w>=\int_{0}^{T}wdt/T\neq 0$, $T$ is the period for the
oscillation of $w$) in this coupled two-component BECs system. When $\left|
K/G\right| <1$, and $\gamma /K$ goes through the bifurcation points which
satisfy $(G/K)^{2/3}-(\gamma /K)^{2/3}=1$, the spin polarization of either
the equal-phase mode or the anti-phase mode is discontinuous at the
bifurcation points. This means a first-order phase transition occurs. For
the zero effective detuning $\gamma $, two metastable states with negative
and positive spontaneous spin polarization coexist. As $\left| K/G\right| $
is increased to $1$, the spontaneous spin polarization vanishes: this
corresponds to a second-order phase transition.

Similar to the thermal atoms, the spontaneous spin polarization can be
induced by adjusting the coupling lasers. Additionally, the collisions among
ultracold atoms can also be controlled. Below, we will consider the
bifurcation and spontaneous spin polarization in Bose condensed atoms
induced by the ultracold collisions. Tuning the coupling laser with fixed
intensity to a certain detuning satisfying $\gamma =0$, the bifurcation and
the spontaneous spin polarization caused by the ultracold collisions in the
above system can be obtained. For the equal-phase mode, only one stable
fixed point $w=0$ exists when $G/K<1$, two new stable fixed points $w_{\pm
}=\pm \sqrt{1-(G/K)^{-2}}$ appear and the original one $w=0$ becomes
unstable when $G/K>1$. This means a Hopf bifurcation occurs at $G/K=1$. The
system goes from the Rabi regime $(G/K<1)$ into the spontaneous spin
polarization regime $(G/K>1)$ through the Hopf bifurcation. However, for the
anti-phase mode, the Hopf bifurcation occurs at $G/K=-1$, there is only one
stable fixed point $w=0$ for $G/K>-1$, there two table fixed points $w_{\pm
}=\pm \sqrt{1-(G/K)^{-2}}$ and an unstable one $w=0$ for $G/K<-1$. The Hopf
bifurcations of both the equal-phase mode and the anti-phase mode are shown
in Fig. 2. The solid lines are stable equilibria (stationary states), the
dot lines are unstable equilibria.

The Hopf bifurcations obtained from analyzing the equilibria in the previous
are static bifurcations. Imagine now that the parameters are swept through
the static bifurcation points. An interesting phenomenon emerges: the system
starting close to the initially stable equilibrium does not immediately
react to the bifurcation. Furthermore, it remains for some time close to the
unstable equilibrium, then fast falls into one of the newly formed stable
equilibria. This has been named as bifurcation delay which has been found in
a variety of physical systems\cite{bfdelay}. The bifurcation delay, which
might lead to hysteresis, is the response to the bistability. For the
equal-phase mode, fixed the effective detuning $\gamma =0$, slowly sweeping
up the ratio $G/K$ from $R_{0}$ with sweeping rate $r$ $($i.e., $%
G/K=R_{0}+rt $, $1\gg r>0)$, choosing $R_{0}<1$ and the initial state close
to the equilibrium, the system evolves along the unstable equilibrium for a
period of time after the ratio sweeping through the static bifurcation point 
$(G/K=1)$, then it quickly goes into a small oscillations around one of two
new stable equilibria. The equilibrium, which the system evolves around
lastly, determines by the state at the static bifurcation point. The system
evolves around the up branch lastly when this state is close to the up
branch; otherwise, the system evolves around the down branch. When $R_{0}>1$%
, slowly sweeping down the ratio through the static bifurcation point with
initial state close to one of two stable equilibria, the system evolves near
the stable equilibrium before it sweeps through the static bifurcation
point, then it goes into a small Rabi oscillation around the ordinary
equilibrium ($w=0$). For the same sweeping rate, averaging the small
oscillations, the process of sweeping up and down generates a hysteresis
loop in the plane extended by $G/K$ and $w$. The area enclosed in the
hysteresis loop increases with the sweeping rate. This means that the energy
exchanged between the atoms and the environments increases with the sweeping
rate. The bifurcation delay in the equal-phase mode with different sweeping
rate is shown in Fig. 3. For the anti-phase mode, a similar behavior can be
observed near the static bifurcation point $G/K=-1$.

To observe the spontaneous spin polarization and the bifurcation delay
predicted in the above, we suggest a experiment based upon the works of JILA%
\cite{Hall}. Two BECs in the $|F=1,m_{F}=-1>$ and $|2,1>$ spin states of $%
^{87}R_{b}$ are coupled by a two-photon pulse with the two-photon
Rabi-frequency $\Omega $ and a finite detuning $\delta $. Thus, the control
of the parameters $K$ and $\gamma $ can be realized by adjusting the Rabi
frequency and the detuning of the coupling lasers, respectively. The tuning
of the parameter $G$ can be accomplished with the Feshbach resonance\cite
{Inouye}. The longitudinal and transverse spin components can be measured
with the state-selective absorption imagining and the Ramsey interference,
respectively\cite{spinwave}.

Summary, due to the coherent ultracold collision among condensed bosonic
atoms, the bifurcation and the spontaneous spin polarization in coupled
two-component BECs are determined by both the relative phase and the
parameters. These phenomena are different from those only determined by the
parameters, we name them as conditional bifurcation and conditional
spontaneous spin polarization, respectively. For the zero effective detuning 
$\gamma $, the Hopf bifurcation and bifurcation delay can be induced by the
Feshbach resonance in either the equal-phase mode or the anti-phase mode.
The system falls into the spontaneous spin polarization regime from the Rabi
regime after the bifurcations occur. The appearance of bifurcation delay
indicates the existence of metastability and hysteresis. Because of the
inherently quantum coherence and superposition of two components, this
quantum hysteresis might open the door to storage quantum data with Bose
condensed atoms\cite{quanthys}.

\begin{center}
{\Large Acknowledgement}
\end{center}

\begin{quotation}
The authors are grateful to the help of Dr. J. Feng in our institute. The
work is supported by the National Natural Science Foundations of China (No.
10275023, 19904013 and 19904014) and foundations of the Chinese Academy of
Sciences.
\end{quotation}

\begin{center}
{\Large Figure capt}{\large ion}
\end{center}

\begin{quotation}

Fig. 1 The fixed points for the system with different ratios $\gamma /K$ and 
$G/K$. The numbers labelled on the lines are values for $G/K$.

Fig. 2 The static Hopf bifurcation and the spontaneous spin polarization.

Fig. 3 The bifurcation delay in the equal-phase mode for different values of
sweeping rate which are labelled on the lines.
\end{quotation}

\epsfxsize6.5truein
\epsfbox{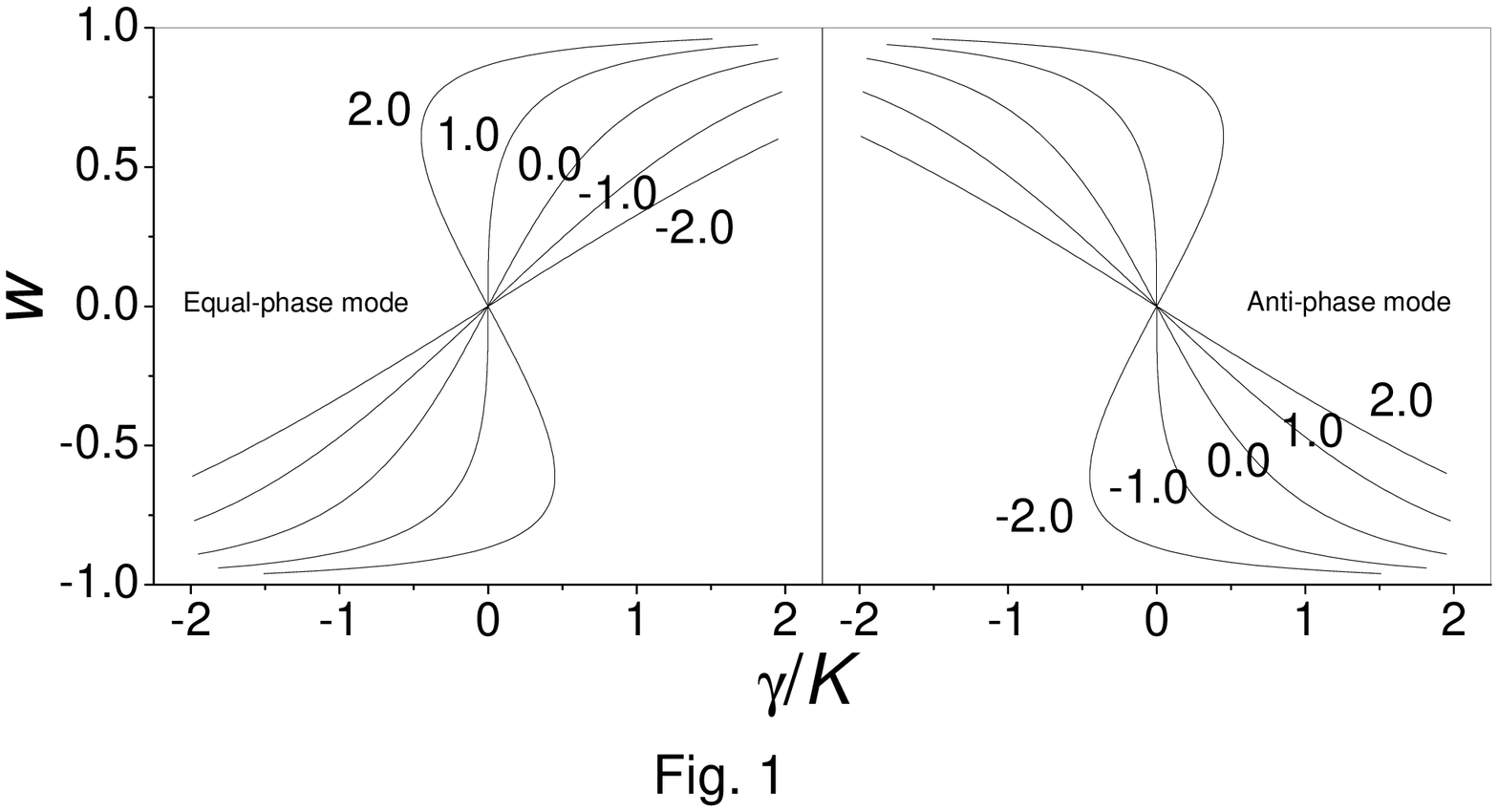}
\epsfxsize6.5truein
\epsfbox{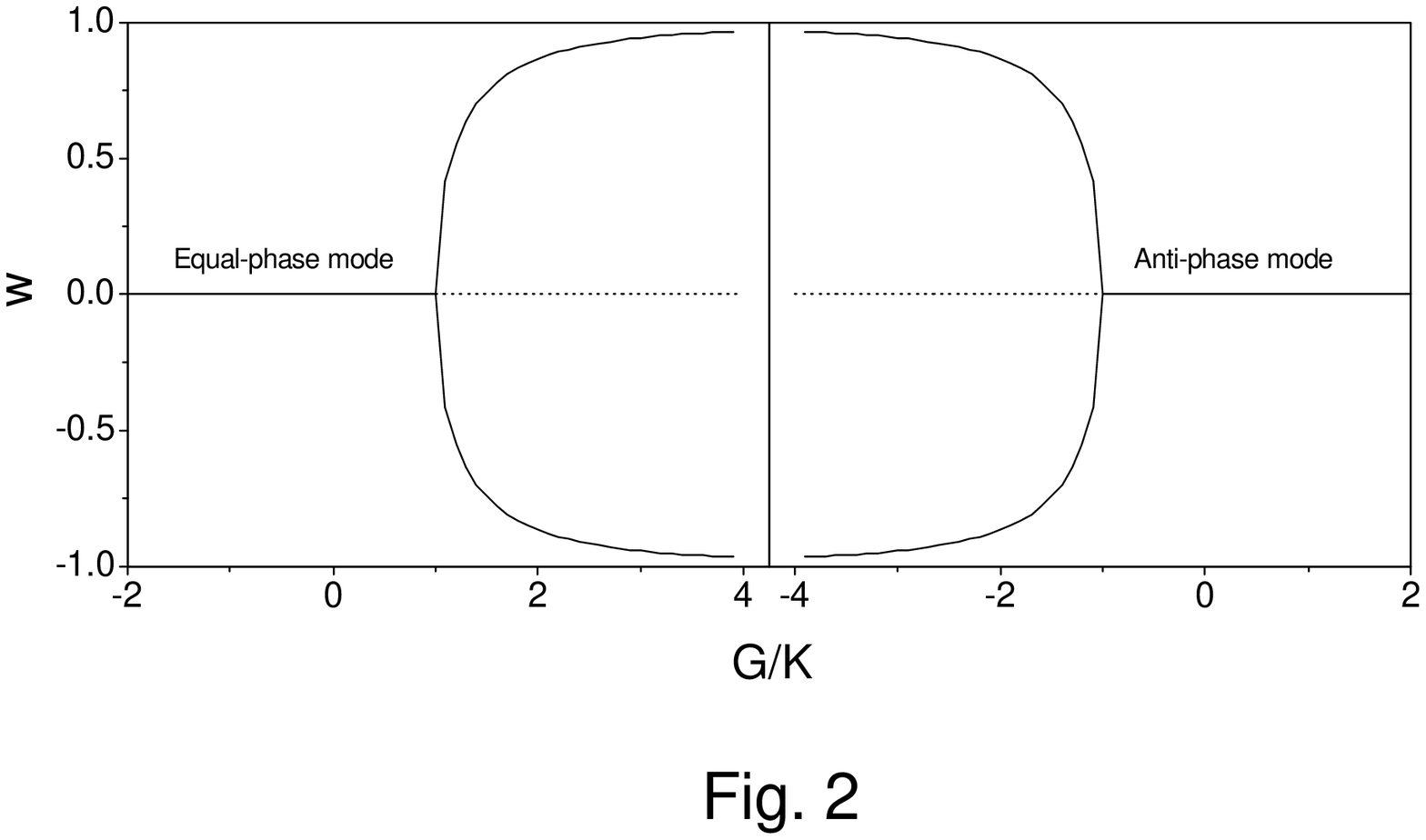}
\epsfxsize6truein
\epsfbox{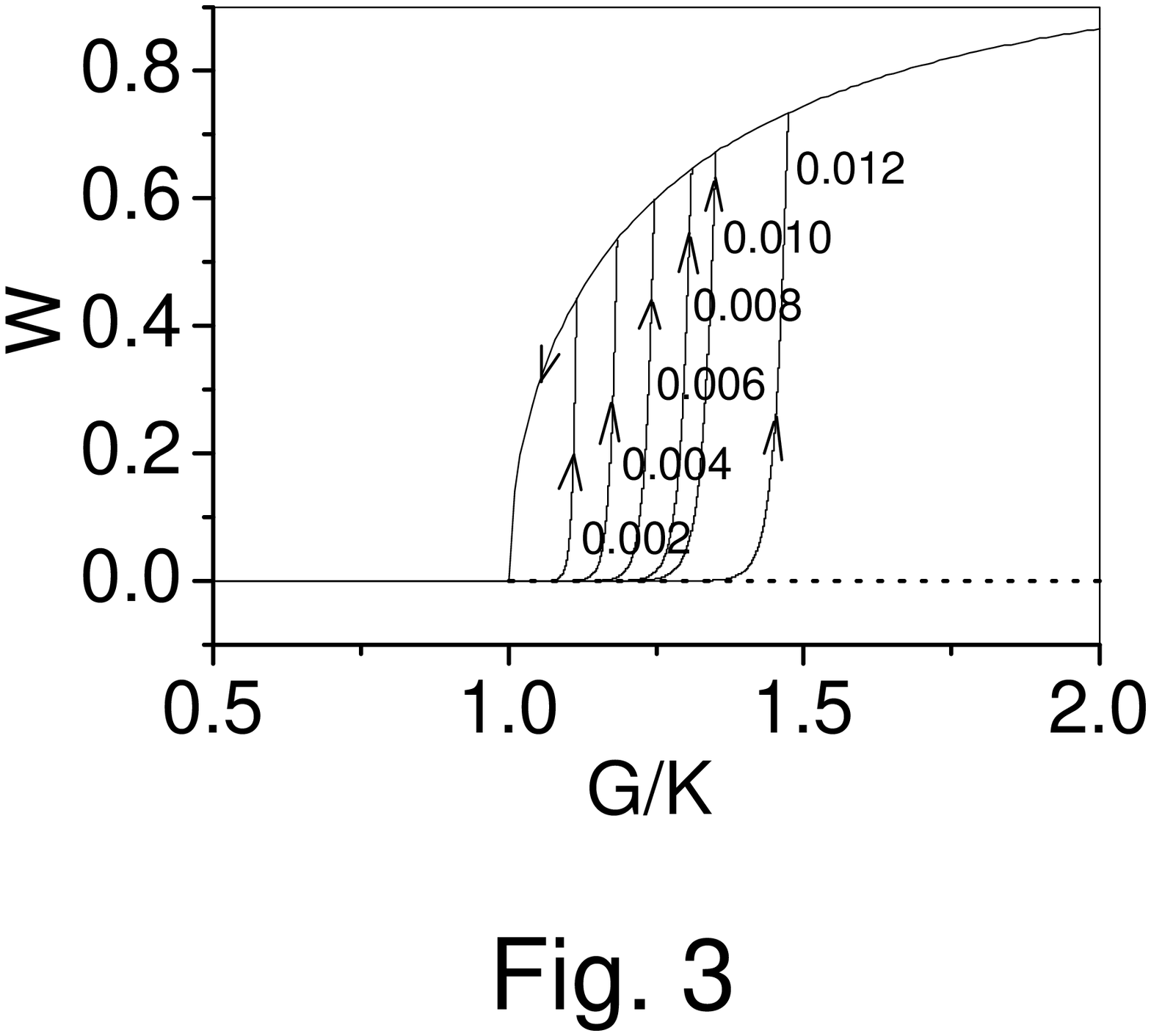}


\begin{references}
\bibitem{Fortson}  N. Fortson and B. Heckel, Phys. Rev. Lett. 59, 1281
(1987); W. M. Klipstein, S. K. Lamoreaux, and E. N. Fortson, Phys. Rev.
Lett. 76, 2266 (1996); A. Andalkar, et. al., Phys. Rev. A 65, 023407 (2002).

\bibitem{Gibbs}  H. M. Gibbs, et al., Phys. Rev. Lett. 36, 113 (1976); T.
Yabuzaki, et al., Phys. Rev. A 29, 1964 (1984).

\bibitem{Rao}  M. Rao, H. R. Krishnamurthy, and R. Pandit, Phys. Rev. B 42,
856 (1990).

\bibitem{Stenger}  J. Stenger, S. Inouye, D. M. Stamper-Kurn, H. -J.
Miesner, A. P. Chikkatur, and W. Ketterle, Nature 396, 345 (1998); C. J.
Myatt, E. A. Burt, R. W. Ghrist, E. A. Cornell, and C. E. Wieman, Phys. Rev.
Lett. 78, 586 (1997).

\bibitem{Hall}  D. S. Hall, M. R. Matthews, J. R. Ensher, C. E. Wieman, and
E. A. Cornell, Phys. Rev. Lett. 81, 1539 (1998); D. S. Hall, M. R. Matthews,
C. E. Wieman, and E. A. Cornell, Phys. Rev. Lett. 81, 1543 (1998).

\bibitem{Burnett}  K. Burnett, P. S. Julienne, P. D. Lett, E. Tiesinga, and
J. Williams, Nature 416, 225 (2002); J. Weiner, V. S. Bagnato, S. Zilio, and
P. S. Julienne, Rev. Mod. Phys. 71, 1 (1999).

\bibitem{Anglin}  J. R. Anglin and W. Ketterle, Nature 416, 211 (2002); F.
Dalfovo, S. Giorgini, L. P. Pitaevskii, and S. Stringari, Rev. Mod. Phys.
71, 463 (1999).

\bibitem{Inouye}  S. Inouye, et. al., Nature 392, 151 (1998); Ph.
Courteille, R. S. Freeland, and D. J. Heinzen, Phys. Rev. Lett. 81, 69
(1998); S. L. Cornish, et. al, Phys. Rev. Lett. 85, 1795 (2000); S. Chu,
Nature 416, 206 (2002).

\bibitem{Kohler}  S. Kohler and F. Sols, Phys. Rev. Lett. 89, 060403 (2002).

\bibitem{Williams}  J. Williams, et al., Phys. Rev. A 59, R31 (1999); J.
Williams, Ph.D thesis, JILA and University of Colorado (1999).

\bibitem{Lee}  C. Lee, W. Hai, L. Shi, X. Zhu, and K. Gao, Phys. Rev. A 64,
053604 (2001); W. Hai, C. Lee, G. Chong, and L. Shi, Phys. Rev. E 66, 026202
(2002);{\bf \ }C. Lee, W. Hai, X. Luo, L. Shi, and K. Gao, arXiv:
cond-mat/0206134 (2002).

\bibitem{Kuang}  L. -M. Kuang and Z. -W. Ouyang, Phys. Rev. A 61, 023604
(2000); B. Hu and L. -M. Kuang, Phys. Rev. A 62, 023610 (2000); L. -M.
Kuang, Z. -Y. Tong, Z. -W. Ouyang, and H. -S. Zeng, Phys. Rev. A 61, 013608
(2000); L. -M. Kuang, J. -H. Li and B. Hu, J. Opt. B: Quantum Semiclass.
Opt. 4, 295 (2002).

\bibitem{coupling}  R. J. Ballagh, K. Burnett, and T. F. Scott, Phys. Rev.
Lett. 78, 1607 (1997); M. $\stackrel{..}{O}$. Oktel and L. S. Levitov, Phys.
Rev. Lett. 83, 6 (1999); Y. Wu and X. Yang, Phys. Rev. A 62, 013603 (2000).

\bibitem{Zoller}  K. Marzlin, W. Zhang, and E. M. Wright, Phys. Rev. Lett.
79, 4728 (1997); R. Dum, J. I. Cirac, M. Lewenstein, and P. Zoller, Phys.
Rev. Lett. 80, 2972 (1998).

\bibitem{spinwave}  H. J. Lewandowski , et al., Phys. Rev. Lett. 88, 070403
(2002); M. $\stackrel{\text{..}}{\text{O}}$. Oktel and L. S. Levitov, Phys.
Rev. Lett. 88, 230403 (2002); J. N. Fuchs, D. M. Gangardt, and F. Lalo$%
\stackrel{\text{..}}{\text{e}}$, Phys. Rev. Lett. 88, 230404 (2002); J. E.
Williams, T. Nikuni, and C. W. Clark, Phys. Rev. Lett. 88, 230405 (2002); J.
M. McGuirk, et al., Phys. Rev. Lett. 89, 090402 (2002); T. Nikuni, J. E.
Williams, and C. W. Clark, Phys. Rev. A 66, 043411 (2002).

\bibitem{doublewell}  M. R. Andrews, et al., Science 275, 637 (1997); C.
Orzel, et al., Science 291, 2386 (2001); A. Smerzi, et al., Phys. Rev. Lett.
79, 4950 (1997); S. Raghavan, et al., Phys. Rev. A 59, 620 (1999); I.
Marino, et al., Phys. Rev. A 60, 487 (1999); A. Smerzi and S. Raghavan,
Phys. Rev. A 61, 063601 (2000).

\bibitem{bfdelay}  P. Mandel and T. Erneux, Phys. Rev. Lett. 53, 1818
(1984); N. Berglund and H. Kunz, Phys. Rev. Lett. 78, 1691 (1997); N.
Berglund and H. Kunz, J. Phys. A 32, 15 (1999).

\bibitem{quanthys}  E. M. Chudnovsky, Science 274, 938 (1996); P. C. E.
Stamp, Nature 383, 125 (1996).
\end{references}
\end{document}